\documentclass{PoS}
\usepackage[utf8]{inputenc}
\usepackage{fancybox}
\usepackage{hhline}
\usepackage{dcolumn}
\usepackage{textcomp}
\usepackage{epsfig,graphics,graphicx}
\usepackage{amsfonts,amssymb,amsmath}
\usepackage{pifont}
\usepackage{bm}
\usepackage{lscape}
\usepackage[mathscr]{euscript}
\usepackage{mathrsfs}
\usepackage{multirow}
\usepackage{lipsum}
\usepackage{color}   

\newcommand{\beq}{\begin{equation}}
\newcommand{\eeq}{\end{equation}}
\newcommand{\bea}{\begin{eqnarray}}
\newcommand{\eea}{\end{eqnarray}}

\newcommand{\ord}[1]{{\cal{O}}( #1 )}
\newcommand{\B}{{\bf B}}
\newcommand{\Bdag}{{\bf B^\dagger}}

\title{Nucleon and Delta axial-vector couplings in\\ $\mathbf{ 1/N_c}$ - Baryon Chiral Perturbation Theory}

\ShortTitle{Axial-vector couplings in $1/N_c$ - ChPT}

\author{
        \speaker{A. Calle Cord\'on}
        \thanks{This work was supported by DOE Contract No. DE-AC05-06OR23177 under which JSA operates the Thomas Jefferson National Accelerator Facility, and by the National Science Foundation (USA) through grant PHY-0855789 (JLG).}\\
        Thomas Jefferson National Accelerator Facility, Newport News, Virginia 23606, USA. \\
        E-mail: \email{cordon@jlab.org}
        }

\author{
        J.~L.~Goity\\
        Thomas Jefferson National Accelerator Facility, Newport News, Virginia 23606, USA.\\
        Department of Physics, Hampton University, Hampton, VA 23668, USA.\\
        E-mail: \email{goity@jlab.org}}

\abstract{ 
In this contribution, baryon axial-vector couplings are studied  in the framework of the combined  $1/N_c$ and chiral expansions~\cite{CalleCordon:2012xz}. This framework is implemented on the basis of the emergent spin-flavor symmetry in baryons at large $N_c$ and HBChPT, and  linking   both expansions ($\xi$-expansion), where $1/N_c$ is taken to be a quantity $\ord{p}$. The study is carried out including one-loop contributions, which corresponds to $\ord{\xi^3}$ for baryon masses and $\ord{\xi^2}$ for the axial couplings.
An analysis of the Lattice QCD results for the axial couplings of both $N$ and $\Delta$ is presented.
}

\FullConference{The 7th International Workshop on Chiral Dynamics,\\
		August 6 -10, 2012\\
		Jefferson Lab, Newport News, Virginia, USA}

\begin{document}

\section{Introduction}
In addition to  approximate chiral $SU_L(N_f)\times SU_R(N_f)$ symmetry in the light quark sector, QCD presents an approximate $SU(2 N_f)$ spin-flavor dynamical symmetry in the baryon sector, which becomes exact in the large $N_c$   and degenerate quark masses limits~\cite{Gervais:1983wq,Dashen:1993as}. An effective theory of baryons which combines these symmetries seems a natural framework,  as it has been advanced in Ref.~\cite{Jenkins:1995gc} and more recently in Ref.~\cite{CalleCordon:2012xz}. The spin-flavor symmetry requires that the effective theory contains ground state baryons, in the symmetric  $SU(2N_f)$  representation with $N_c$ indices, with spins ranging from 1/2 to $N_c/2$. For two flavors and $N_c=3$  the symmetry is $SU(4)$ and the states are the $N$ and $\Delta$ baryons. The spin-flavor symmetry plays a key role in controlling the loop contributions in the effective theory: while the coupling of pions to baryons diverge as $\sqrt{N_c}$, in loop contributions there are cancellations which keep the effective theory consistent with a natural $1/N_c$ power counting~\cite{Witten:1979kh}. While loop contributions to baryon masses diverge as $N_c$ in the case of the spin-flavor singlet component of the masses,  the spin-flavor breaking effects are   $\ord{1/N_c}$. Similarly, the loop contributions to currents, e.g., the axial currents discussed here, respect the power counting as a result of strict cancellations \cite{CalleCordon:2012xz,FloresMendieta:2000mz}.
One expects that any quantity in baryons, where spin-flavor symmetry imposes cancellations, will be more naturally described if the effective theory is built in accordance with the dictates of that symmetry. This in particular avoids the need for introducing large counterterms, as it occurs in BChPT involving only the nucleons. The progress of baryon lattice QCD (LQCD) results for masses and other observables \cite{CD12talks}  is creating  new avenues  for understanding low energy baryon physics,  giving access to the quark mass dependence of those observables, and also to the $N_c$ dependencies as advanced in   the recent work \cite{DeGrand:2012hd}. These developments will in particular help establish the most useful baryon effective theory, from its very framework  to the values of the low energy constants (LECs) involved. 
This contribution outlines the description of the axial couplings of $N$ and $\Delta$ in the combined $1/N_c$ and HBChPT framework in the $\xi$-expansion scheme of Ref.~\cite{CalleCordon:2012xz}, and presents and analysis in that framework of LQCD results for the axial couplings $g_A^{NN}$, $g_A^{N\Delta}$, and $g_A^{\Delta\Delta}$.

\section{Axial-vector couplings in  $1/N_c$  HBChPT}

In two-flavor QCD, the isovector axial current is given by:
\bea
A_\mu^a = \bar{q}\ \gamma_\mu \gamma_5 \frac{\tau^a}{2}\ q\, ,
\eea
where $\tau^a$ are  Pauli isospin matrices.
Based on symmetries,
the matrix elements of the axial current between baryon states are parametrized in terms of form factors~\cite{Adler:1968tw,Adler:1975mt,Llewellyn Smith:1971zm}. The axial coupling is given by the value of the form factor associated with the piece of the axial current that  dominates in the limit of vanishing three-momentum transfer, and is obtained in that limit, and in the baryon rest frame, from the matrix elements of the spatial components of the axial current.  Terms other than the axial coupling one are suppressed by a factor $\Delta m/m_B=\ord{1/N_c^2}$, with $\Delta m$  the mass difference between the initial and final  baryons. For $N$ and $\Delta$ one therefore has:
\bea
\left\langle N(p,s')\vert A_i^a \vert N(p,s)\right\rangle &=&G_A(q^2=0)\, \bar u(p,s') \, \gamma_i \, \gamma_5  \,
  \frac{\tau^a}{2} \,  u(p,s) \, ,\\
\label{eq:NN-ME}
 \left\langle \Delta^+(p,s')\vert A_i^3 \vert p (p,s)\right\rangle &=&  \sqrt{\frac{2}{3}}\,   C_5^A(q^2 = 0)\,
\bar u_{\Delta^+}^\mu (p,s')\, g_{i\mu}\,  u_p(p,s)\, ,\\
\label{eq:ND-ME}
\left\langle \Delta^+(p,s')\vert A_i^3 \vert \Delta^+(p,s)\right\rangle &=& \frac{1}{2}\, g_1(q^2=0) \,
\bar u_{\Delta^+}^\mu (p,s')\,
g_{\mu\nu}\,
\gamma_i\, \gamma_5\,
u_{\Delta^+}^\nu (p,s)\, ,
\label{eq:DD-ME}
\eea
where $q_\mu = (p' - p)_\mu$ is the momentum transfer, $u(p,s)$ is the Dirac spinor corresponding to the nucleon, and $u_\mu$ is the Rarita-Schwinger vector-spinor corresponding to the $\Delta$.
The nucleon's axial coupling is defined as $g_A^{NN} = G_A(0)$ with the experimental value $1.2701 \pm 0.0025$~\cite{Beringer:1900zz}.
In Eq. (2.3) $p(p,s)$ indicates the proton, and $C_A^5(q^2)$ is the analogue of the nucleon axial form factor $G_A(q^2)$. Finally, the axial coupling of the $\Delta$ is given by $g_1(0)$. 


The chiral Lagrangian for the combined  $1/N_c$  HBChPT  in the $\xi$-expansion, which is defined by linking the chiral and $1/N_c$ expansions according to  $1/N_c=\ord{p}$, reads at leading order~\cite{CalleCordon:2012xz}:
\bea
{\cal{L}}^{(1)}_\B &=&\Bdag\left(i\, D_0 +  \mathring{g}_A
 u^{ia}G^{ia} - \frac{C_{HF}}{N_c}{ \vec{S}^2}-\frac{c_1}{2} N_c\; \chi_+\right)\B\, .
\label{eq:Lagrangian-LO}
\eea
Here $\bf B$ represents an $SU(4)$ multiplet in the symmetric representation,  which consists of $N$ and $\Delta$ when $N_c=3$. $G^{ia}$ are spin-flavor generators, $\vec{S}$  are  the spin generators, and the rest are the usual building blocks in HBChPT.

In Ref.~\cite{CalleCordon:2012xz}, baryon masses and axial couplings were evaluated up to $\ord{\xi^3}$ and $\ord{\xi^2}$ respectively. That evaluation involves  one-loop contributions, which are spelled out in all detail in that reference. In general, the $\xi$ power counting of a Feynman diagram with a single baryon line flowing through it, is determined by the general formula~\cite{CalleCordon:2012xz}:
\beq
\nu_\xi=1+3 L+\frac{n_\pi}{2} +\sum_i n_i\;(\nu_{O_i}+\nu_{p_i}-1) \, ,
\label{eq:xi-counting}
\eeq
where $L$ is the number of loops, $n_\pi$ the number of external pions,   $i$ indicates the type of vertex, $n_i$ the number of such  vertices in the diagram, $\nu_{p_i}$ the chiral order of the vertex,  and  $\nu_{O_i}$ the $1/N_c$ order   of the spin-flavor operator in the vertex ~\footnote{
The $1/N_c$ order $\nu_O$ of the spin-flavor operator $O$ is given by $\nu_O = n - 1 - \kappa$, where $n$ is the number of generators appearing as factors in the operator (we say that the operator is $n$-body) and $\kappa$ is the number of generators $G^{ia}$ that appear in the product (see Ref.~\cite{Dashen:1994qi} for further details). 
}.

\begin{center}
\begin{figure*}[ttt]
\centerline{
\includegraphics[width=3.5cm,angle=0]{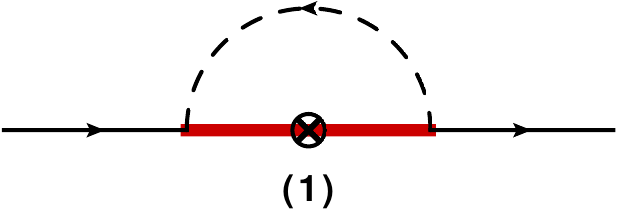}
\includegraphics[width=3.5cm,angle=0]{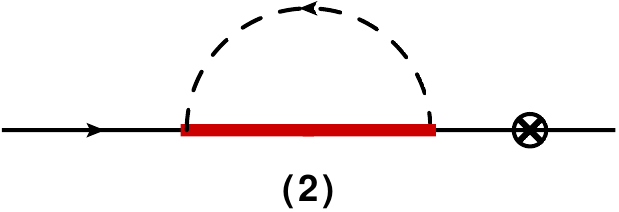}
\includegraphics[width=3.5cm,angle=0]{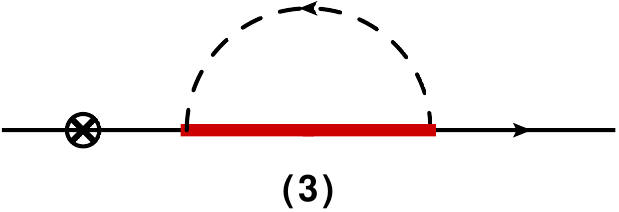}
}
\caption{
One loop corrections to the axial-currents in the $\xi$-expansion  contributing up to $\ord{\xi^2}$. 
The crossed circle  denotes the axial-current operator. }
\label{fig:loop-vertex}
\end{figure*}
\end{center}

The counterterm Lagrangians needed for renormalizing the one-loop evaluation of masses and axial couplings are given by:
\bea
{\cal{L}}_{\Sigma}^{CT}&=&\Bdag\left\{\frac{m_1}{N_c} +\frac{C_{HF1}}{N_c^2}\vec{S}^2+\frac{C_{HF2}}{N_c^3}\vec{S}^4 \right.+\mu_1 \chi_+ +\frac{\mu_2}{N_c} \chi_+ \vec{S}^2  \nonumber \\
& &\hspace*{.8cm} +\ \left. \left(\frac{w_1}{N_c}+\frac{w_2}{N_c}\vec{S}^2+\frac{w_3}{N_c^3}\vec{S}^4+(z_1 N_c+\frac{z_2}{N_c}\vec{S}^2)\chi_+\right)(iD_0-\delta m)\right\}\B,\\
\label{eq:self-energy-CT-lagrangian}
{\cal{L}}{  _{A}^{CT}}&=&\Bdag u^{ia}\left( \frac{C^A_0}{N_c} G^{ia}+\frac{C^A_1}{4}\{\chi_+,G^{ia}\}+ \frac{C^A_2}{N_c^2} \{\vec{S}^2,G^{ia}\}+ \frac{C^A_3}{N_c} [\vec{S}^2,G^{ia}]+
\frac{C^A_4}{N_c} S^iI^a
\right)\B~,
\label{eq:axial-current-CT-lagrangian}
\eea
where the LECs are determined in the MS scheme in the usual way. As defined here, all LECs are $\ord{N_c^0}$, and of course sub leading $1/N_c$ corrections are implicit in them. One notable result is the contribution to the wave function renormalization, which requires the counterterm proportional to $z_1$: this CT is $\ord{N_c p^3}$.  Such behavior is however essential for cancelling with  terms in the evaluation of the one-loop corrections to the axial couplings which violate $N_c$ power counting. On the other hand, it evidently shows that taking the limit $N_c\to \infty$ makes the chiral expansion ill defined. This is however not a surprise, since the two expansions do not commute, thus the necessity to define a linking between them as it is done here with the $\xi$-expansion. The diagrams in Fig.~\ref{fig:loop-vertex} give the one-loop corrections to the axial current, their contributions to the axial couplings being of both $\ord{\xi}$ and $\ord{\xi^2}$. 

At lowest order, the spatial components of the axial currents are simply given by:
$A^{ia} =  \mathring{g}_A\ G^{ia} $. Upon including corrections, their matrix elements in the  $q\to 0$ limit are expressed by:
 \beq
 \langle\B'\mid A^{ia}\mid\B\rangle=g_A^{\B\B'} \; \frac{6}{5}\;{ \langle \B' \mid G^{ia} \mid \B \rangle} \, .
\label{eq:gA-formula}
 \eeq
 Here $g_A^{BB'}$   corresponds for $N_c=3$  to the standard definition of the nucleon axial coupling,   the factor $6/5$ being included for that reason  \footnote{In Eq. (21) of Ref.~\cite{CalleCordon:2012xz} there is a mistake, limited to that Eq., where the factor is indicated, incorrectly, as $5/6$.}. The axial couplings defined by Eqs.~\eqref{eq:NN-ME} to~\eqref{eq:DD-ME}, which are given in the LQCD results  analyzed below,  are related to the $g_A^{BB'}$ as follows:
 \bea
 g_A^{NN} = G_A(0) \equiv g_A \, , \hspace*{0.3cm} 
 g_A^{N\Delta} = \frac{5}{3\sqrt{2}} \ C_5^A(0)\, , \hspace*{0.3cm}
 g_A^{\Delta\Delta} = \frac{5}{3} \  g_1(0)\, .
\eea
The axial couplings defined here are $\ord{N_c^0}$ and the $\ord{N_c}$ of the matrix elements of the axial currents stems from  the operator $G^{ia}$. 

\section{Confronting  LQCD results}

In this section the results for the three axial couplings obtained in LQCD calculations are analyzed~\footnote{For LQCD results in baryons, and in particular the nucleon's axial coupling, see Ref.~\cite{CD12talks}}. Of particular interest is   the quark mass, or pion mass, dependence of those couplings. At  present,  there are first results at different values of $N_c$ only for the baryon masses, and thus the analysis at fixed $N_c=3$ cannot fix the $N_c$ sub leading dependencies of the LECs; in particular this means that the LEC $C_0^A$ in Eq.~\eqref{eq:axial-current-CT-lagrangian} cannot be determined, and either $C_4^A$, which for $N_c=3$ multiplies a term which is linearly dependent with the terms proportional to  $C_2^A$ and  $C_3^A$. Therefore in the fit below $C_0^A$ and $C_4^A$ are set to vanish.
The analysis presented here involve the simultaneous fitting to baryon masses and axial couplings as in Ref.~\cite{CalleCordon:2012xz}, but this time including the axial couplings $g_A^{N\Delta}$ and $g_A^{\Delta\Delta}$. The LQCD results are from the ETM Collaboration: for $g_A^{NN}$ Ref.~\cite{Alexandrou:2010hf},  and  for $C_5^A(0)$, $g_1(0)$, and the $N$ and $\Delta$ masses  the results are those in Refs.~\cite{Alexandrou:2011py} and~\cite{Alexandrou:private}.

\begin{centering}
\begin{table}[ttt]
\caption{The LECs in the table correspond to the choice of the renormalization scale $\mu=700$ MeV. The baryon masses extrapolated to the physical point are $m_N = 962(15)$ MeV, and  $m_\Delta = 1213(14)$ MeV, and the axial couplings extrapolated to the physical point are $g_A = 1.17(2)$, $ C_5^A(0)= 0.91(2)$ and $ g_1(0) = 0.59(2)$.}
\resizebox{1.\textwidth}{!}
{
\begin{tabular}{ c c c c c c c c c c }
\hline\hline
$\chi^2_{\rm DOF}$ & $\mathring{g}_A$ & $m_0$ & $C_{\rm HF}$ & $c_1$ & $\mu_2$ & $z_1$ & $C_1^A$ & $C_2^A$ & $C_3^A$\\
 & & [MeV] & [MeV] & [MeV$^{-1}$] & [MeV$^{-1}$] & [MeV$^{-2}$] & [MeV$^{-2}$] &  & \\
[1mm]\hline ~\\ [-.5mm]
$ 2.0$ & $1.55(3)$ & $261(8)$ & $174(11)$ & $0.0025(1)$ & $-0.0003(2)$ & $-1.0(1)\times 10^{-6}$ & $-6(2)\times 10^{-7}$ & $-0.24(5)$ & $-0.15(8)$ \\ [1.5mm]\hline\hline
\end{tabular}
}
\label{tab:lattice}
\end{table}
\end{centering}

\begin{figure*}[ttt]
\begin{center}
\includegraphics[width=7cm,height=5cm,angle=0]{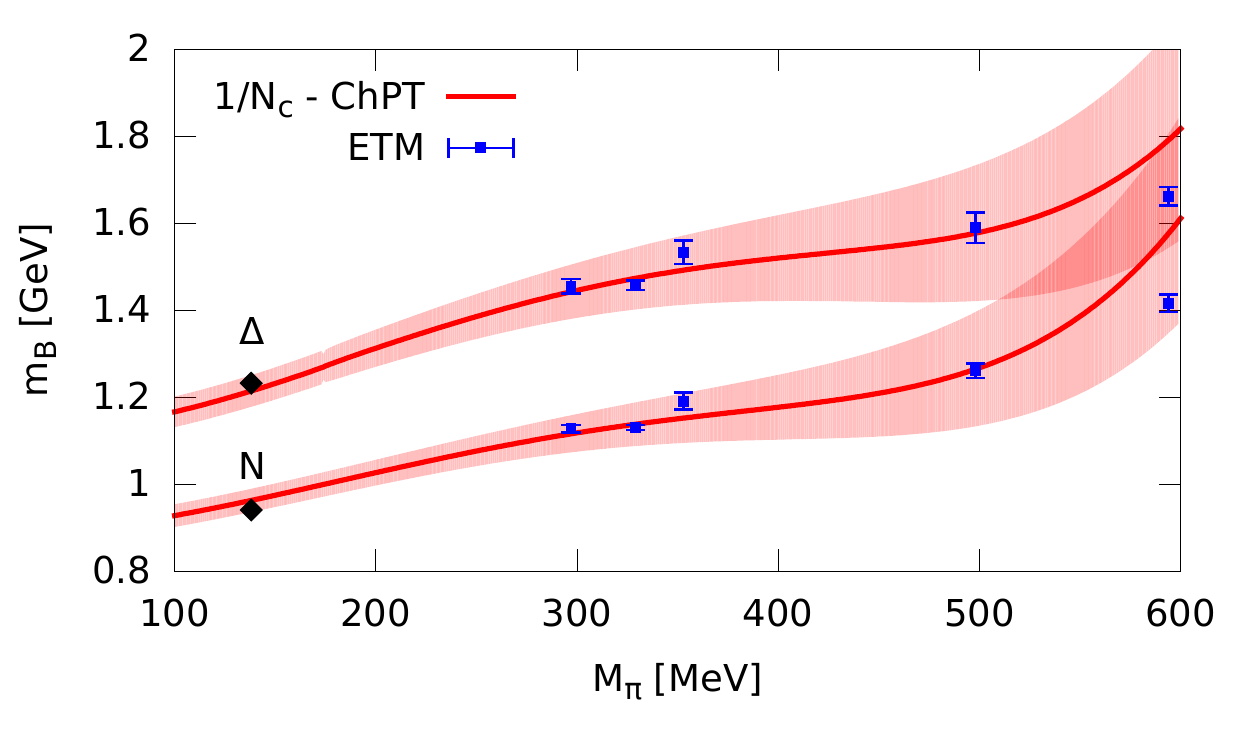}
\includegraphics[width=7cm,height=5cm,angle=0]{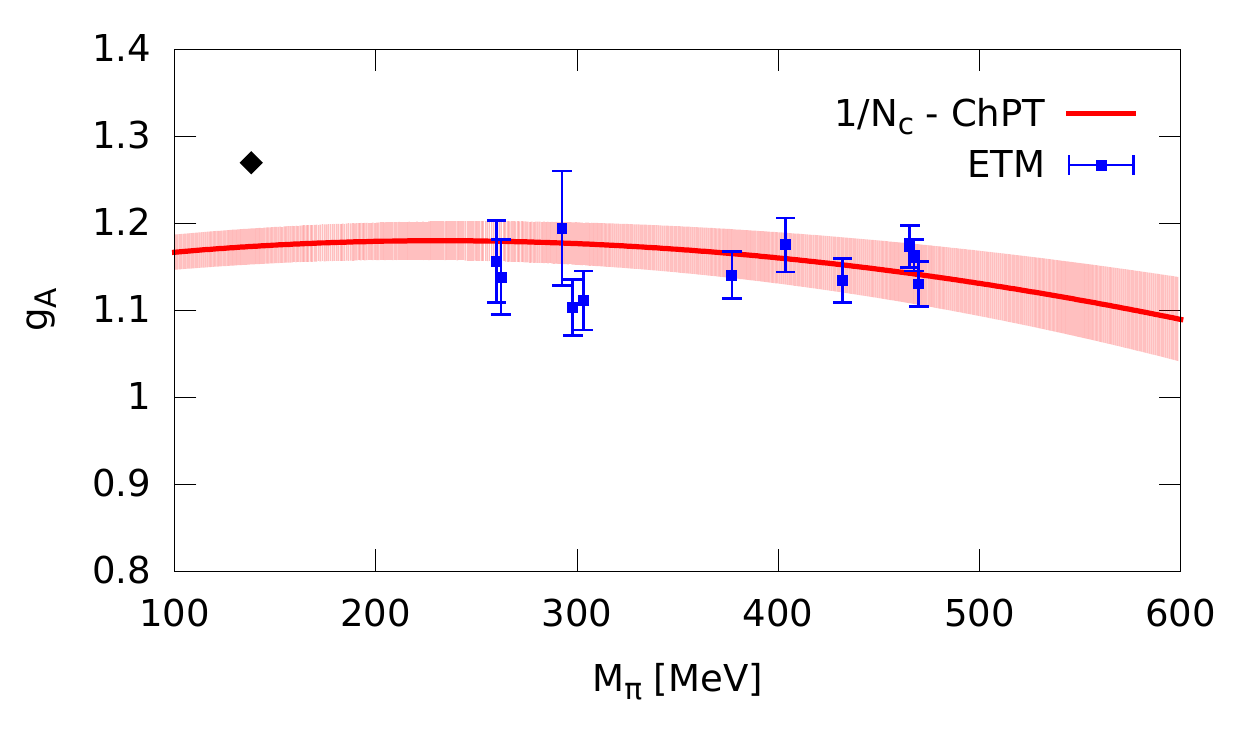}\\
\includegraphics[width=7cm,height=5cm,angle=0]{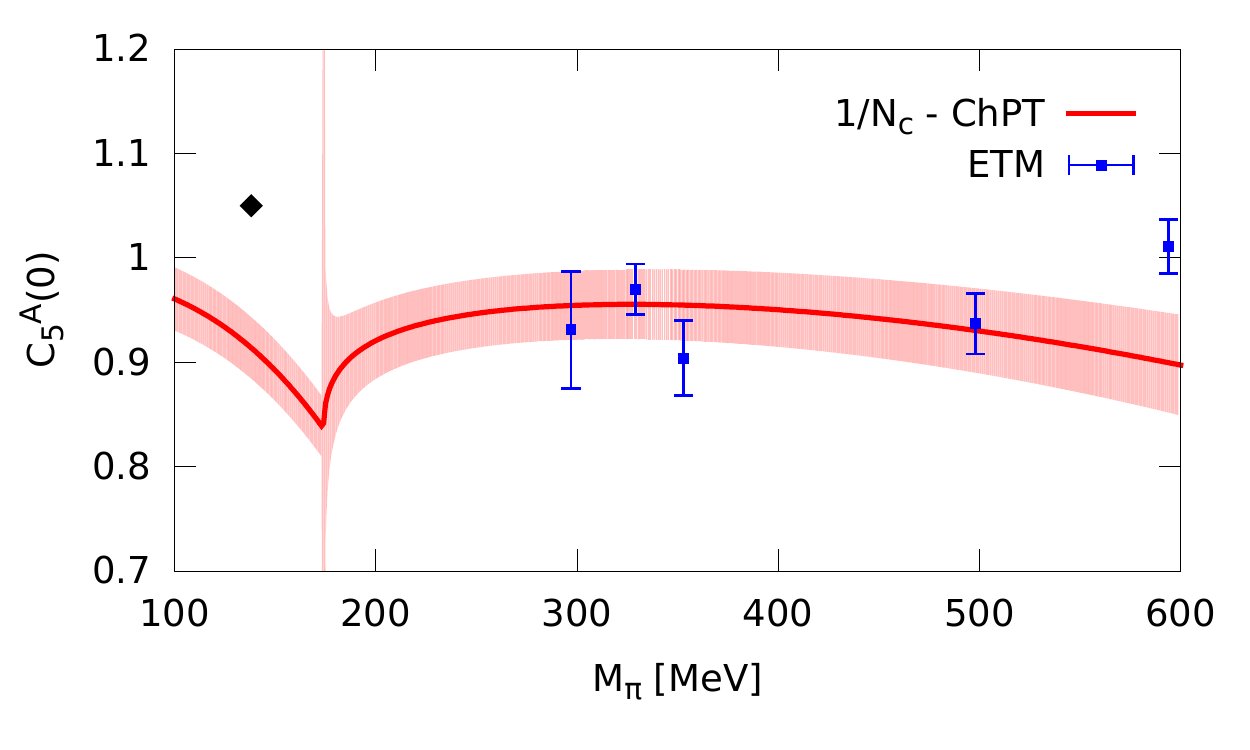}
\includegraphics[width=7cm,height=5cm,angle=0]{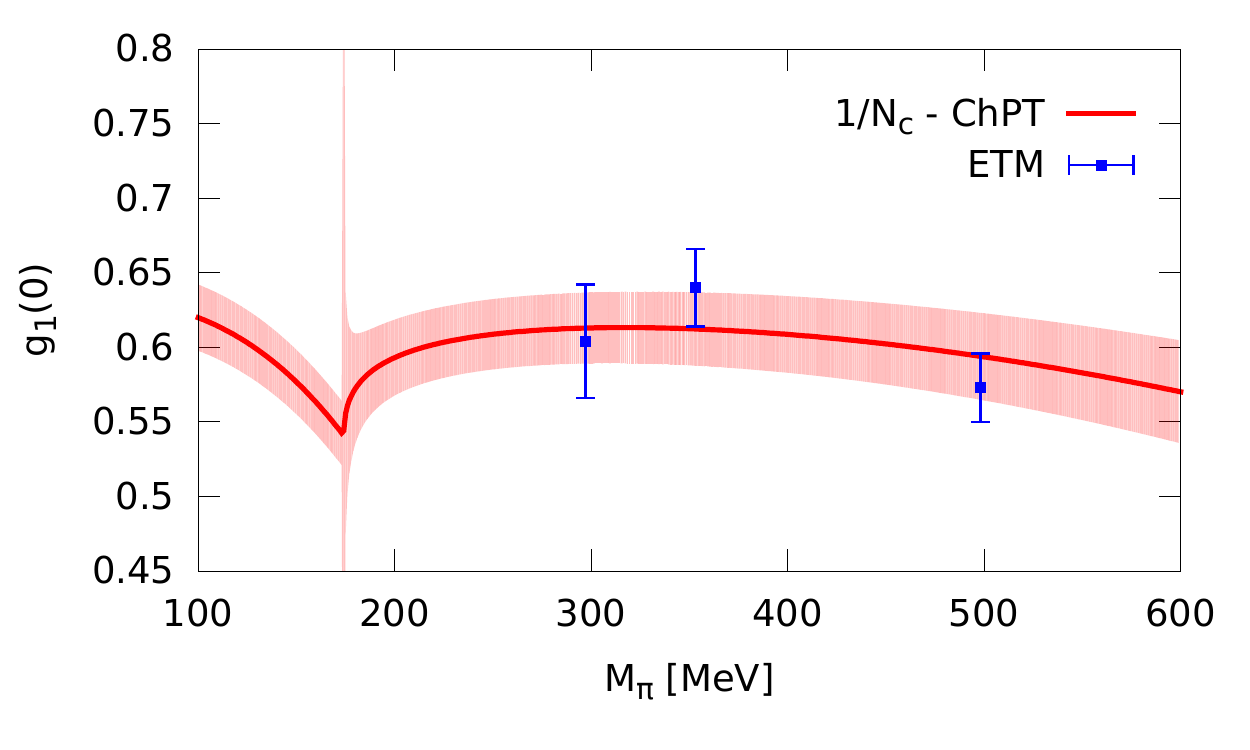}
\end{center}
\caption{Combined at NNLO to the masses and axial coupling of the ETM collaboration~\cite{Alexandrou:2010hf,Alexandrou:2011py,Alexandrou:private}. The diamonds depict the physical values. The bands correspond to the theoretical 68\% confidence interval.}
\label{fig:lattice}
\end{figure*}

The combined  fit to the $N$ and $\Delta$ masses at $\ord{\xi^3}$  and the axial couplings at  $\ord{\xi^2}$ as functions of $M_\pi$ are carried out for the available data below  $M_\pi\sim 500$ MeV. The results obtained are displayed in Table~\ref{tab:lattice} and in Fig.~\ref{fig:lattice}.
The $\Delta$ width is used to determine the physical value of $C_5^A$ according to: $\Gamma_{\Delta\to N\pi}=\frac{C_5^{A^2}}{6\pi F_\pi^2}((m_N-m_\Delta)^2-M_\pi^2)^{3/2}$, which upon using $ \Gamma_{\Delta\to N\pi}^{\rm Exp}=116-120$ MeV, gives $C_5^A\sim 1.05\pm 0.01$. As emphasized in~\cite{CalleCordon:2012xz} this corresponds vis-\`a-vis the nucleon axial coupling to a remarkably small deviation from the $SU(4)$ symmetry limit. One can see that the LQCD results show, for the $\Delta N$ axial coupling, a similar deficit when extrapolated to the physical point as it occurs for the $N$ axial coupling. The cusp observed in the axial couplings involving the $\Delta$ are due to the opening of the $\Delta\to N\pi$ channel. Its location is not the physical one because the baryon masses in the loop are the ones at $\ord{\xi}$.

The rather flat behavior of the lattice results for the axial couplings within the mass range considered here is reasonably  reproduced by the effective theory; the LECs turn out to have natural size. There is however some curvature in the effective theory, resulting mostly from the non-analytic contributions, which will require more accurate LQCD results to be tested.  As it was discussed in Ref.~\cite{CalleCordon:2012xz} at length, the small dependence with $M_\pi$  can only be reproduced thanks to the cancellations dictated by the spin-flavor symmetry between different one-loop contributions.
The LECs $C_2^A$, $C_3^A$   give spin-flavor breaking contributions to the axial couplings. Their natural size is 1, but they are significantly smaller as a result of the fit, which gives indication of the spin-flavor breaking in the axial couplings to be suppressed dynamically, as it was mentioned earlier.

As it is known from all LQCD extrapolations to the physical pion mass, $g_A^{NN}$ is underestimated  by about 10\%, and a similar underestimation occurs with $C_5^A$. This is an open issue which in Ref.~\cite{CalleCordon:2012xz} is argued to be of some systematic LQCD origin, and which should be clarified  in forthcoming LQCD calculations.

\section{Conclusions}
The $1/N_c$ expansion at the hadronic level has many claims to fame. It provides an additional bookkeeping tool which is rigorously rooted in QCD, and its blending with effective theories, in particular ChPT, represents the ultimate paradigm for the description of the strong interactions at the hadronic level. It is particularly important in baryons, where the improvement over ordinary BChPT  including only the  spin-1/2 baryon degrees of freedom   is  evident. The applications to masses and axial couplings have been worked out~\cite{CalleCordon:2012xz,FloresMendieta:2006ei,FloresMendieta:2012dn}, and confronted with LQCD results  as  discussed here. Further applications are very promising, so stay tuned.

The authors thank C. Alexandrou for useful communications regarding the ETM results.

\end{document}